\documentclass[12pt]{article}
\usepackage{graphicx}
\usepackage[utf8]{inputenc}
\usepackage{mathtools}
\usepackage{graphicx,psfrag,epsf,color}
\usepackage{amsmath,amssymb,amsfonts}
\usepackage{array}
\usepackage{cite}
\usepackage{float}
\usepackage{enumitem}
\usepackage{appendix}
\usepackage{bm}
\usepackage{ascii}
\usepackage{dsfont}
\usepackage{caption}
\usepackage{subcaption}
\usepackage[colorlinks]{hyperref}
\usepackage[dvipsnames]{xcolor}
\hypersetup{pageanchor=false,linkcolor=Maroon,citecolor=MidnightBlue,urlcolor=MidnightBlue}

\renewcommand{\Im}{\textrm{Im}}
\renewcommand{\vec}[1]{\bm{#1}}
\newcommand{\cor}[2]{#1\langle #2 #1\rangle}
\newcommand{\p}{\partial}
\newcommand{\vp}{\varphi}
\newcommand{\ve}{\varepsilon}
\newcommand{\der}{\textrm{d}}
\newcommand{\Lo}{\mathcal{O}}

\newcommand{\Z}{\mathbb{Z}}

\newcommand{\tmu}{\tilde\mu}

\newcommand{\secref}[1]{Sect.~\ref{#1}}
\newcommand{\figref}[1]{Fig.~\ref{#1}}

\renewcommand{\log}{\ln}

\newcounter{ASQ}

\newcounter{MCFQ}

\numberwithin{equation}{section}

\begin{document}
\allowdisplaybreaks

\begin{titlepage}

\vskip1cm

\begin{center}
{\Large \bf Classical conformal invariance and\\ superhorizon dynamics in de Sitter}
\end{center}

  \vspace{0.5cm}
\begin{center}
{\sc Maria Cristina Fiore and Andrea F. Sanfilippo} 
\\[6mm]
{\it Departamento de F\'isica Te\'orica y del Cosmos, Universidad de Granada,}\\
{\it Campus de Fuentenueva, E–18071 Granada, Spain}
\end{center}
\vskip1cm

\begin{abstract}
\noindent 
Soft de Sitter Effective Theory is a well-motivated candidate for the correct effective late-time description of equal-time correlation functions in de Sitter space. In this work, we study its application to theories that enjoy classical conformal invariance, using the conformally-coupled $\phi^4$-theory as a toy model. While quantum effects generate non-trivial late-time dynamics in such models, we argue that it is not described by the standard construction of the effective theory as discussed thus far in the literature. We show that the tree-level matching of the trispectrum onto the effective theory does not fit into the expected power-counting scheme, and we contrast it with the matching of the tree-level bispectrum in the conformally-coupled $\phi^3$-theory, where it works consistently. We then propose a prescription to identify the leading superhorizon degrees of freedom in such theories, which should serve as the starting point for the construction of their late-time effective description. The interpretation of logarithms of the form $\log(-k\eta)$ in this context is briefly discussed.
\end{abstract}

\end{titlepage}

\section{Introduction}

Soft de Sitter Effective Theory (SdSET) \cite{Cohen:2020php,Cohen:2021fzf} is a viable candidate for the correct effective-field-theory (EFT) framework describing the superhorizon dynamics of quantum fields in de Sitter (dS) space. 
So far, it has been applied primarily to the case of the massless, minimally coupled, real scalar field with a quartic self-interaction \cite{Cohen:2020php,Cohen:2021fzf,Beneke:2026rtf,Beneke:2026ksj}. In that model, it is the natural tool to tame the infrared divergences that plague the perturbative expansion of correlation functions, allowing one to recover, and systematically extend, the formalism of stochastic inflation first derived in the pioneering works \cite{Starobinsky:1982ee,Starobinsky:1986fx,Starobinsky:1994bd}. Further applications include the re-derivation \cite{Cohen:2020php} of the conservation of the scalar and tensor modes of the metric perturbation outside the horizon during inflation \cite{Maldacena:2002vr,Weinberg:2003sw,Weinberg:2003ur,Assassi:2012et}, the study of the rare tails of probability distributions relevant to eternal inflation \cite{Cohen:2021jbo}, the description of the superhorizon dynamics of tensor perturbations in inflationary spacetime \cite{Green:2024fsz}, and the study of principal-series \cite{Cohen:2024anu} and compact scalars \cite{Chakraborty:2025mhh} in dS.

In this work, we continue the exploration of the structure and applications of the SdSET by considering theories that enjoy classical conformal invariance, using the quartically self-interacting, conformally-coupled (CC) real scalar field as a paradigmatic example.  The tree-level correlation functions of this theory are equivalent, up to an overall dependence on the scale factor, to flat-space correlation functions of a massless scalar field with a quartic self-interaction. As such, the classical theory is insensitive to the presence of the horizon of dS, and no redshifting of the modes takes place. However, the conformal invariance is broken at loop level by ultraviolet (UV) divergences, since the renormalization procedure introduces a dependence of the correlation functions on the renormalization scale. The superhorizon dynamics of the quantized theory is thus non-trivial, but it is entirely generated by quantum effects. In \cite{Green:2020txs} it was argued that, as a consequence of this, the deviation of the functional form of correlators in such theories from their flat-space counterpart is directly linked to the logarithms of the renormalization scale, and can thus be controlled using the standard renormalization-group (RG) equations. This fact was then exploited to realize the idea of the ``dynamical renormalization group" (DRG) \cite{Tanaka:1975ti,Boyanovsky:1998aa,Boyanovsky:2003ui,Burgess:2009bs} and to study the effect of the resummation of secular logarithms in this class of theories. While advantageous from this point of view, the loop-induced sensitivity to the dynamical spacetime also makes the identification of the superhorizon degrees of freedom, and the construction of their appropriate effective description, subtle. The CC $\phi^4$-theory has been discussed in the context of the SdSET in \cite{Cohen:2020php}, but this point seems to have gone unnoticed.  

The CC scalar field with a quartic self-interaction is often used in the literature on cosmological correlators as a toy model due to its simplicity, and it is especially popular for the study of the structure of loop diagrams in dS, see \cite{Green:2020txs,Heckelbacher:2022fbx,Lee:2023jby,Benincasa:2024ptf,Chowdhury:2025ohm,Nowinski:2025cvw} for an incomplete list. Its classical conformal invariance is shared by more realistic theories, such as four-dimensional (non-)Abelian gauge theories with massless fermions. Although correlation functions of both gauge fields and spin-$1/2$ fermions are suppressed by inverse powers of the scale factor in the late-time limit \cite{Baumann:2020dch}, it is still interesting to understand what their effective description in this limit would entail, which motivates this work.

The paper is structured in the following way. We introduce the models of interest in \secref{sec:CCtheory}, and give the general argument for the tree-level equivalence of the $\phi^4$-theory to a flat-space theory. We introduce the CC SdSET in \secref{sec:CCSdSET}, building on the formalism set up in \cite{Cohen:2020php,Beneke:2026rtf}, and we first use it for the tree-level matching of the bispectrum in the CC $\phi^3$-theory in \secref{sec:bimatch}. The conformal invariance in this theory is broken already at tree level, and, as a consequence of this, the matching succeeds. We then contrast this with the case of the CC $\phi^4$-theory in \secref{sec:trimatch}. We attempt to match the tree-level trispectrum onto SdSET, and show that the necessary terms do not fit into the expected power-counting scheme. We identify the leading superhorizon behaviour of the two-point function of the interacting field in \secref{sec:superhdofs}, which arises from a two-loop effect, and argue that this result can be used to identify the proper superhorizon degrees of freedom. We conclude in \secref{sec:conclusion}.

\section{Conformally-coupled scalar field in de Sitter}
\label{sec:CCtheory}

To set the stage, we introduce our notation and conventions. We work with the dS metric in flat slicing, using the signature $(+,-,...,-)$. The line element reads
\begin{equation}
\der s^2=g_{\mu\nu}(x)\der x^{\mu}\der x^{\nu}=a(\eta)^2\Big[\der\eta^2-\der\vec x^2\Big]\,,
\end{equation}
where we use the conformal-time variable $\eta\in(-\infty,0)$, the scale factor 
\begin{equation}
a(\eta)=-\frac{1}{H\eta}\,,
\end{equation}
and the constant Hubble parameter $H$. We will also make use of the cosmological-time coordinate $t\in(-\infty,\infty)$, which measures the ticks of a clock of a comoving observer. It is related to $\eta$ through
\begin{equation}
    \der t=a(\eta)\der\eta\,,
\end{equation}
and the scale factor as a function of $t$ takes the exponential form $a(t)=e^{Ht}$. The class of theories we consider is defined by the free action
\begin{equation}
S=\frac{1}{2}\int\der^dx\;\sqrt{-g}\Big[g^{\mu\nu}\p_{\mu}\phi\p_{\nu}\phi-\Big(m^2+\xi R\Big)\phi^2\Big]\,,
\label{eq:freeS}
\end{equation}
with the Riemann scalar $R=d(d-1)H^2$. We work in the Bunch-Davies vacuum \cite{Bunch:1978yq}, in which the mode functions take the form
\begin{equation}
\phi_{\vec k}(\eta)=-ie^{\frac{i\pi}{2}(\nu+\frac{1}{2})}\sqrt{\frac{\pi}{4H}}(-H\eta)^{\frac{d-1}{2}}H^{(1)}_{\nu}(-k\eta)\,,
\label{eq:modefct}
\end{equation}
with $H^{(1)}_{\nu}(z)$ the Hankel function of the first kind. The parameter $\nu$ is defined as
\begin{equation}
\nu\equiv\sqrt{\bigg(\frac{d-1}{2}\bigg)^2-\frac{m^2}{H^2}-\xi d(d-1)}\,.
\end{equation}
The constant phase factor in \eqref{eq:modefct} is conventional, and drops out when computing correlation functions of the field $\phi$. We keep the dimensionality of spacetime $d$ general, in order to use dimensional regularization (dimreg) \cite{tHooft:1972tcz,Bollini:1972bi} to regularize the UV divergences encountered in loop diagrams. 

Following \cite{Melville:2021lst} we define the conformally-coupled free field in $d$ dimensions by picking
\begin{equation}
m^2=0\,,\quad \xi=\frac{d-2}{4(d-1)}\,,
\label{eq:ddimCC}
\end{equation}
which ensures that $\nu=1/2$ in any dimension. As is well known, with this choice of parameters the free theory can be mapped to the theory of a free massless scalar field $\chi$ in Minkowski spacetime. This may be seen explicitly by means of the field redefinition
\begin{equation}
\phi(\eta,\vec x)=a(\eta)^{\frac{2-d}{2}}\chi(\eta,\vec x)\,.
\label{eq:phichi}
\end{equation}
Accordingly, the Bunch-Davies mode function for $\chi$ reads
\begin{flalign}
\chi_{\vec k}(\eta)=-i\frac{e^{-ik\eta}}{\sqrt{2k}}\,,
\end{flalign}
which has the form of a massless flat-space mode function multiplied by a constant (irrelevant) phase. This result reflects the fact that the irreducible representation of the de Sitter group corresponding to the conformally-coupled scalar field maps smoothly in the limit $H\to0$ to the irreducible representation of the Poincar\'e group corresponding to the massless scalar field \cite{Garidi:2003ys}.

Upon turning on the interactions, the preservation of the equivalence of this theory to a flat-space one depends on the details of the interaction terms. In this work, we consider the contact interactions
\begin{equation}
S_{\textrm{int}}=-\int\der^dx\;\sqrt{-g}\frac{\kappa_n}{n!}\phi^n
\end{equation}
with $n=3,4$. The total action expressed in terms of $\chi$ reads
\begin{equation}
S=\int\der^dx\;\bigg[\frac{1}{2}\p_{\mu}\chi\p^{\mu}\chi-\frac{\kappa_n}{n!}a(\eta)^{d(1-\frac{n}{2})+n}\chi^n\bigg]\,,
\label{eq:chiS}
\end{equation}
where in the above equation the indices are raised and lowered with the Minkowski metric $\eta_{\mu\nu}$. The exponent of $a(\eta)$ in \eqref{eq:chiS} for $d=4$ and $n=3,4$ reads
\begin{equation}
d\bigg(1-\frac{n}{2}\bigg)+n=\begin{cases}
1\,,& n=3\,,\\
0\,,& n=4\,.
\end{cases}
\label{eq:expcases}
\end{equation}
In the former case the interaction is relevant, and the effective interaction strength of the field grows as $\eta\rightarrow0$. On the other hand, for the latter case the classical action \eqref{eq:chiS} is fully independent of the scale factor. This makes the equivalence of the tree-level dynamics of the theory to its flat-space counterpart manifest, despite being formulated in a dynamical spacetime.

The role of observables in the present setting is played by equal-time correlation functions. To compute them we work in the Schwinger-Keldysh \cite{Schwinger:1960qe,Keldysh:1964ud} formalism, using the notation set up in Appendix A of \cite{Beneke:2023wmt}, and we work in spatial momentum space. The correlation functions of $\phi$ and $\chi$ are related to all orders in perturbation theory by the relation
\begin{equation}
\cor{}{\phi(\eta,\vec k_1)...\phi(\eta,\vec k_n)}=a(\eta)^{n\frac{2-d}{2}}\cor{}{\chi(\eta,\vec k_1)...\chi(\eta,\vec k_n)}\,.
\end{equation}
The basic building block for their perturbative evaluation is the Wightman function
\begin{equation}
\cor{}{\phi(\eta,\vec k)\phi(\eta',-\vec k)}'=(-H\eta)^{\frac{d-2}{2}}(-H\eta')^{\frac{d-2}{2}}\frac{e^{ik(\eta'-\eta)}}{2k}\,,
\label{eq:freeCC2pt}
\end{equation}
where the prime denotes that we stripped off an overall momentum-conserving delta function. The values of the exponent of $a(\eta)$ for the two cases \eqref{eq:expcases} imply that the correlation functions of the two theories have a completely different behaviour as $\eta\to0$, which impacts the structure of their effective description in this limit. In the following, we show how this distinction manifests itself when attempting to match the simplest non-trivial correlation functions in the two interacting theories onto SdSET.

\section{Conformally-coupled SdSET}
\label{sec:CCSdSET}

In this section we collect the elements of the SdSET for $\nu = 1/2$ which will be needed in the following discussion. We rely heavily on the formalism and notation set up in \cite{Cohen:2020php,Beneke:2026rtf}, and in the SdSET it is convenient to use the cosmological time variable $t$ instead of $\eta$.

The starting point for its construction is the decomposition of the free scalar field
\begin{equation}
\phi(t,\vec x)=H^{\frac{d}{2}-1}\Big[(a(t)H)^{-\alpha}\vp_+(t,\vec x)+(a(t)H)^{-\beta}\vp_-(t,\vec x)\Big]\,,
\label{eq:freephisplit}
\end{equation}
which is appropriate in the late-time limit\footnote{Notice that this field decomposition is not equivalent to the redefinition \eqref{eq:phichi}, which can be used for any value of $t$ or $\eta$.}, where for $\nu=1/2$ we have
\begin{equation}
\alpha=\frac{d-2}{2}\,,\quad \beta=\frac{d}{2}\,.
\label{eq:CCpowercount}
\end{equation}
Defining the SdSET power-counting parameter
\begin{equation}
\lambda\sim\frac{k}{a(t)H}\ll 1
\end{equation}
the free effective fields $\vp_{\pm}$ are assigned the following counting:
\begin{equation}
\vp_+(t,\vec x)\sim\lambda^{\alpha}\,,\quad\vp_-(t,\vec x)\sim\lambda^{\beta}\,.
\end{equation}
They satisfy the canonical equal-time commutation relations
\begin{flalign}
[\vp_+(t,\vec x),\vp_-(t,\vec y)]&=-i\delta^{(d-1)}(\vec x-\vec y)\label{eq:cancom}\,,\\
[\vp_+(t,\vec x),\vp_+(t,\vec y)]&=[\vp_-(t,\vec x),\vp_-(t,\vec y)]=0\,.
\end{flalign}
Their two-point functions in momentum space are time-independent at leading power, and read
\begin{flalign}
\cor{}{\vp_+(t,\vec k)\vp_+(t',-\vec k)}'&=\frac{1}{2k}\,,\label{eq:vppp2pt}\\
\cor{}{\vp_{\pm}(t,\vec k)\vp_{\mp}(t',-\vec k)}'&=\mp\frac{i}{2}\,,\label{eq:vppm2pt}\\
\cor{}{\vp_-(t,\vec k)\vp_-(t',-\vec k)}'&=\frac{k}{2}\,.
\end{flalign}

Moving on to the interacting CC SdSET, its action reads
\begin{equation}
S[\vp_{\pm}]=S_{\textrm{kin}}[\vp_{\pm}]+S_{\textrm{int}}[\vp_{\pm}]\,,
\end{equation}
with
\begin{flalign}
S_{\textrm{kin}}[\vp_{\pm}]&=-\frac{1}{2}\int\der^{d-1}x\der t\;\Big[\dot\vp_+\vp_--\vp_+\dot\vp_-\Big]\,,\\
S_{\textrm{int}}[\vp_{\pm}]&=-\int\der^{d-1}x\der t\;\sum_{n=1}^{\infty}\sum_{m=0}^na(t)^{3-n-m+(n-2)\ve}\frac{c^0_{n-m,m}}{(n-m)!m!}\vp^{n-m}_+(t,\vec x)\vp_-^m(t,\vec x)\,.\label{eq:Sint}
\end{flalign}
We used $d=4-2\ve$, and the exponent of $a(t)$ in \eqref{eq:Sint} ensures the invariance of the action under dilatations and special conformal transformations \cite{Beneke:2026rtf}. Above, and in the following, the superscript ``0" denotes bare quantities. We restrict ourselves to working only at leading power in the gradient terms, which are thus omitted above, and the effective fields require no renormalization at leading power in $\lambda$. The bare effective couplings $c^0_{n-m,m}$ are related by the reparametrization invariance (RPI) of the SdSET as
\begin{equation}
c^0_{n-m+1,m-1}=Hc^0_{n-m,m}\,.
\label{eq:coupRPI}
\end{equation}

The action is complemented by the initial-condition (IC) functional
\begin{flalign}
\mathcal{F}[\vp_{\pm}]&=\sum_{\sigma=\pm}\sum_{n=1}^{\infty}\int\bigg[\prod_{j=1}^n\frac{\der^{d-1}k_j}{(2\pi)^{d-1}}\bigg]\;\sum_{m=0}^n\frac{a^{3-n-m+(n-2)\ve}_*}{(n-m)!m!}\Xi^{\sigma;0}_{n-m,m}(\vec k_1,...,\vec k_n)\nonumber\\
&\phantom{=}\times(2\pi)^{d-1}\delta^{(d-1)}\bigg(\sum_{l=1}^n\vec k_l\bigg)\prod_{l=m+1}^{n}\vp^{\sigma}_+(t_*,\vec k_l)\prod_{q=1}^m\vp^{\sigma}_-(t_*,\vec k_q)\,.
\label{eq:FIC}
\end{flalign}
Here $a_* \equiv a(t_*)$ is the scale factor evaluated at a reference time value $t_*$, which should be physically thought of as being close to the time of horizon crossing of the effective modes. However, as it plays the role of one of the two factorization scales of the SdSET, it can take any value of $a$ within the regime of validity of the EFT \cite{Beneke:2026rtf}. The bare IC functions $\Xi^{\sigma;0}_{n-m,m}$ encode the evolution of the EFT modes before horizon crossing. They are restricted by the Hermiticity of the initial-state density matrix that they parametrize to satisfy \cite{Garny:2009ni}
\begin{equation}
i\Xi^{\sigma;0}_{n-m,m}(\vec k_1,...,\vec k_n)=\Big[i\Xi^{-\sigma;0}_{n-m,m}(\vec k_1,...,\vec k_n)\Big]^*\,.
\label{eq:IChermitean}
\end{equation}
As was done in \cite{Beneke:2026rtf}, we already restricted the form of the IC functional to contain only terms which do not mix the two branches of the Schwinger-Keldysh path integral. However, as we will see below, the additional relation $\Xi^{+;0}_{n-m,m}=-\Xi^{-;0}_{n-m,m}$ which was implemented in that reference proves too restrictive in this setting. There, it was assumed that the IC functions $\Xi^{\sigma;0}_{n-m,m}$ themselves should be real-valued, which is however not necessary, see e.g. \cite{Agarwal:2012mq}. At this point, we will leave the general CTP-index structure open, and constrain it in the matching step. Since equal-time in-in correlation functions of real fields are real-valued, these objects do not contain enough information to fix the real and imaginary parts of the $\Xi^{\sigma;0}_{n-m,m}$ independently. For our purposes here, it is however sufficient to require the following additional property from the IC functions: They should allow for the matching to work while preserving manifest power-counting at the level of the IC functional. This means that the $\lambda$-scaling of any operator in \eqref{eq:FIC} should be read off from the effective-field content alone, as is the case for the Lagrangian interactions. This enforces $\Xi^{\sigma;0}_{n-m,m}\sim k^0_i\sim\lambda^0$. 

As was pointed out in \cite{Beneke:2026rtf}, the functional \eqref{eq:FIC} contains integer powers of $a_*$. As the full functional must be $a_*$-independent, this implies that the IC functions themselves contain an implicit dependence on integer powers of $a_*$ which compensates for this. This spurious dependence can obscure the power-counting of these terms. To avoid this complication we adopt the prescription proposed in \cite{Beneke:2026rtf} and let
\begin{equation}
a^{3-n-m+(n-2)\ve}_*\to a(t)^{3-n-m}a^{(n-2)\ve}_*
\end{equation}
in \eqref{eq:FIC}, with $t$ the correlation time. Retaining the $\ve$-dependent powers of $a_*$ ensures that we may track any logarithms of $a(t)$ generated by the UV divergences of the SdSET correlators.

Both the Lagrangian and IC terms inherit any discrete symmetry present in the full theory. In the case of the $\phi^4$-theory, the $\Z_2$-symmetry under $\phi\to-\phi$ leads to 
\begin{equation}
c^0_{n-m,m}=0\,,\quad \Xi^{\sigma;0}_{n-m,m}=0\,,\quad n\textrm{ odd}\,.
\end{equation}
In this work we will restrict ourselves to tree-level computations in the SdSET, so we may drop the superscript ``0" on the effective couplings. On the other hand, the IC functions still require renormalization, so we keep their superscript. For a discussion of the renormalization procedure for these objects, see \cite{Beneke:2026rtf}.

As discussed in \cite{Beneke:2026rtf} for the massless, minimally coupled scalar field, the decomposition \eqref{eq:freephisplit} gets corrected when considering the interacting theory. This is so because the dynamical degrees of freedom of the free and interacting SdSET are not the same. One way in which this manifests is the necessity to eliminate Lagrangian interactions $\sim c_{n,0}\vp^n_+$ by means of a field redefinition of $\vp_-$ of the form
\begin{equation}
\vp_-(t,\vec x)\to\vp_-(t,\vec x)+\frac{a(t)^{3-n+(n-2)\ve}}{3-n+(n-2)\ve}\frac{c_{n,0}}{(n-1)!}\vp^{n-1}_+(t,\vec x)
\label{eq:CCredef}
\end{equation}
to arrive at a basis of the effective fields which displays manifest power-counting. The assumption underlying this procedure is that the interacting superhorizon degrees of freedom can be related perturbatively to the free ones appearing in \eqref{eq:freephisplit}. As we will show below, this assumption is not viable in the case of the CC $\phi^4$-theory.

\section{Matching the tree-level bispectrum in \texorpdfstring{$\phi^3$}{phi3}-theory}
\label{sec:bimatch}

As a first application of the formalism set up above we consider the tree-level bispectrum of the CC $\phi^3$-theory introduced in \secref{sec:CCtheory} in $d=4$ dimensions. The conformal invariance of the free theory is broken already at tree level in this example, and we show that the matching may be performed consistently within the power-counting scheme presented in \secref{sec:CCSdSET}.

\subsection{Full-theory computation}

To compute the tree-level bispectrum we need to evaluate
\begin{flalign}
\cor{}{\phi(\eta,\vec k_1)
\phi(\eta,\vec k_2)
\phi(\eta,\vec k_3)}'&=-
\frac{\kappa_3(-H\eta)^3}
{4k_1k_2k_3}\Im\left[\int_{-\infty}^{\eta}\frac{\der\eta'}{(-H\eta')}\,e^{ik_t(\eta-\eta')}
\right]\nonumber\\
&=-\frac{(-H\eta)^3}{4k_1k_2k_3}\frac{\kappa_3}{H}\Im\Big[
e^{ik_t\eta}\Gamma(0,ik_t\eta)\Big]\,,
\label{eq:phi3bisfull}
\end{flalign}
where $k_t=\sum_{i=1}^3k_i$, and $\Gamma(a,z)$ denotes the incomplete Gamma function. Expanding the above result in the limit $-k_t\eta\to0$, we obtain
\begin{flalign}
\lim\limits_{-k_i\eta\rightarrow0}\cor{}{\phi(\eta,\vec k_1)\phi(\eta,\vec k_2)\phi(\eta,\vec k_3)}'
&=\frac{(-H\eta)^3}{4k_1k_2k_3}\frac{\kappa_3}{H}\bigg\{-\frac{\pi}{2}+\Big[1-\log(-e^{\gamma_E}k_t\eta)\Big](-k_t\eta)\nonumber\\
&\phantom{=}+\Lo\Big((-k_t\eta)^2\Big)\bigg\}\,.
\label{eq:latebispectrum}
\end{flalign}
The leading term in the expansion of the integral is consistent with the findings of~\cite{Arkani-Hamed:2015bza}. Notice that, at sub-leading power, a very mild form of secular-logarithmic enhancement appears, which is, however, still damped by the coefficient $(-\eta)^4$. This enhancement may be related to the fact that, in this theory, the strength of the self-interaction of the field effectively grows with time, as can be seen from \eqref{eq:chiS} with $n=3$. 

\subsection{Matching onto SdSET}

To set up the matching computation, we need to establish the matching equation between the full-theory and SdSET correlation functions. Since, in general, SdSET correlators are UV-divergent already at tree level, we work in $d=4-2\ve$ dimensions in the EFT to regulate them. After renormalizing the EFT correlators, we let $\ve\to0$ and match onto the four-dimensional full-theory results. 

The decomposition of the interacting full-theory field in terms of interacting SdSET fields gets modified by the necessity of eliminating the Lagrangian interaction term $\sim c_{3,0}\vp^3_+$ using \eqref{eq:CCredef} for $n=3$:
\begin{equation}
\vp_-(t,\vec x)\rightarrow\vp_-(t,\vec x)+\frac{c_{3,0}a(t)^{\ve}}{2\ve}\vp^2_+(t,\vec x)\,.
\label{eq:cubicvpredef}
\end{equation}
This leads to the matching equation
\begin{flalign}
&\lim\limits_{k_i/(a(t)H)\to 0}\cor{}{\phi(t,\vec k_1)\phi(t,\vec k_2)\phi(t,\vec k_3)}'=\lim\limits_{\ve\rightarrow0}\bigg\{a(t)^{-3+3\ve}\bigg[\cor{}{\vp_+(t,\vec k_1)\vp_+(t,\vec k_2)\vp_+(t,\vec k_3)}'\nonumber\\
&+(a(t)H)^{-1}\bigg(\cor{}{\vp_-(t,\vec k_1)\vp_+(t,\vec k_2)\vp_+(t,\vec k_3)}'+\frac{c_{3,0}a(t)^{\ve}}{2\ve}\cor{}{\vp^2_+(t,\vec k_1)\vp_+(t,\vec k_2)\vp_+(t,\vec k_3)}'\nonumber\\
&+\textrm{permutations}\bigg)\bigg]\bigg\}
\label{eq:bimatch}
\end{flalign}
between full-theory and SdSET three-point functions, up to next-to-leading power in $\lambda$. Our goal is now to use it to reproduce the late-time limit of the full-theory result \eqref{eq:latebispectrum}.

\subsubsection{Leading-power matching}

The leading term in \eqref{eq:latebispectrum} is time-independent, and as such it should be matched by means of an IC insertion into the $\vp_+$ bispectrum. The term in \eqref{eq:FIC} with $n=3$, $m=0$, 
\begin{equation}
\sum_{\sigma=\pm}
\int\bigg[\prod_{j=1}^3\frac{\der^{d-1}k_j}{(2\pi)^{d-1}}\bigg](2\pi)^{d-1}\delta^{(d-1)}\bigg(\sum_{l=1}^3\vec k_l\bigg)\frac{a^{\ve}_*}{3!}\Xi^{\sigma;0}_{3,0}(\vec k_1,\vec k_2,\vec k_3)\vp^\sigma_+(t_*,\vec k_1)\vp^\sigma_+(t_*,\vec k_2)\vp^\sigma_+(t_*,\vec k_3)
\label{eq:F30}
\end{equation}
has the appropriate field content to reproduce the term of interest. Using the free $\vp_+$ two-point function \eqref{eq:vppp2pt} the insertion of $\Xi^{\sigma;0}_{3,0}$ into the $\varphi_+$ bispectrum yields
\begin{equation}
\langle
\vp_+(t,\vec k_1)
\vp_+(t,\vec k_2)
\vp_+(t,\vec k_3)
\rangle'_{|\,\Xi_{3,0}}
=\frac{ia^{\ve}_*}{8k_1k_2k_3}\Big[\Xi^{+;0}_{3,0}(\vec k_1,\vec k_2,\vec k_3)+\Xi^{-;0}_{3,0}(\vec k_1,\vec k_2,\vec k_3)\Big]\,.
\label{eq:F30result}
\end{equation}
Plugging \eqref{eq:F30result} into the matching equation \eqref{eq:bimatch} only fixes the symmetric combination $\Xi^{+;0}_{3,0}+\Xi^{-;0}_{3,0}$. As commented above, we cannot fix $\Xi^{+;0}_{3,0}$ and $\Xi^{-;0}_{3,0}$ individually, but we find that to reproduce the leading term in~\eqref{eq:latebispectrum} it is sufficient to match the purely imaginary, constant and finite IC function 
\begin{equation}
\Xi^{+;0}_{3,0}(\vec k_1,\vec k_2,\vec k_3)=\Xi^{-;0}_{3,0}(\vec k_1,\vec k_2,\vec k_3)=\frac{i\pi\kappa_3}{2H}\,.
\label{eq:Xi30matching}
\end{equation}
We see that the resulting expression is $\sim k^0_i\sim\lambda^0$, as desired, as the operator content of \eqref{eq:F30} correctly captures the necessary power-counting to reproduce the leading expression in~\eqref{eq:latebispectrum}. 

It is also instructive to compare the present situation to the case of the massless, minimally coupled scalar field discussed in \cite{Beneke:2026rtf}. In that case, the IC terms $\sim\Xi^{\sigma;0}_{2n,0}\vp^{2n}_+$ counted as ``super-leading" with respect to the kinetic term of the SdSET. However, due to the CTP-index structure used in that reference for the IC functions, they could only contribute to correlation functions containing a single $\vp_-$-field, and therefore the power-enhancement was spurious.  Nevertheless, they yielded terms at the same order in the power-counting as IC terms $\sim\Xi^{\sigma;0}_{2n-1,1}\vp^{2n-1}_+\vp_-$ inserted into an all-$\vp_+$ correlator. This feature would have obscured the $\lambda$-counting of the EFT correlation functions, and it was found that these terms could be consistently excluded. Here the term $\sim\Xi^{\sigma;0}_{3,0}\vp^3_+$ counts as leading power, and we find that it is both consistent and necessary to include it in the matching. To get a non-vanishing result we had to allow for a more general CTP-index structure than the one considered in \cite{Beneke:2026rtf}. Since the found result for $\Xi^{\sigma;0}_{3,0}$ satisfies the constraint \eqref{eq:IChermitean}, there is no obstruction in doing so.  

\subsubsection{Next-to-leading-power matching}

At next-to-leading power we need to consider both the insertions of the power-suppressed Lagrangian- and IC terms into the $\vp_+$ bispectrum, as well as the insertion of the leading-power IC term found above into the correlation functions in the second line of \eqref{eq:bimatch}. 

We start with the former. After removing the interaction term proportional to $\varphi_+^3$ through the field redefinition~\eqref{eq:cubicvpredef}, the leading-power Lagrangian interaction reads
\begin{equation}
-\int \der^{d-1}x\der t\,
a(t)^{-1+\ve}\frac{c_{2,1}}{2}
\varphi_+^2(t,\vec{x})\varphi_-(t,\vec{x})\,.
\label{eq:cubicLPint}
\end{equation}
Inserting it into the $\vp_+$-bispectrum we find
\begin{flalign}
\cor{}{\vp_+(t,\vec k_1)\vp_+(t,\vec k_2)\vp_+(t,\vec k_3)}'_{|\,c_{2,1}}&=-\frac{ic_{2,1}}{4k_1k_2k_3}\int_{-\infty}^{t}\der t'\;a(t')^{-1+\varepsilon}\sum_{i=1}^{3}k_i
\cor{}{[\vp_+(t,\vec k_i),\vp_-(t',\vec k_i)]}\nonumber\\
&=\frac{c_{2,1}}{4k_1k_2k_3}\frac{k_t}{a(t)}\,,
\label{eq:c21matching}
\end{flalign}
where we used \eqref{eq:cancom} and we already let $\ve\to0$, since the time integral is finite. We also need to consider the insertion of the next-to-leading order IC term
\begin{flalign}
&\sum_{\sigma=\pm}\int\bigg[\prod_{j=1}^3\frac{\der^{d-1}k_j}{(2\pi)^{d-1}}\bigg](2\pi)^{d-1}\delta^{(d-1)}\bigg(\sum_{l=1}^3\vec k_l\bigg)\frac{a(t)^{-1}a^{\ve}_*}{2}\Xi^{\sigma;0}_{2,1}(\vec k_1,\vec k_2,\vec k_3)\vp^{\sigma}_+(t_*,\vec k_1)\vp^{\sigma}_+(t_*,\vec k_2)\nonumber\\
&\times\vp^{\sigma}_-(t_*,\vec k_3)
\end{flalign}
into the $\vp_+$-bispectrum, which yields
\begin{flalign}
\cor{}{\vp_+(t,\vec k_1)\vp_+(t,\vec k_2)\vp_+(t,\vec k_3)}'_{|\,\Xi_{2,1}}
&=
\frac{ia(t)^{-1}a_*^{\varepsilon}}{4k_1k_2k_3}
\sum_{j=1}^3k_j\Big[\Xi^{+;0}_{2,1}(\vec k_1,\vec k_2,\vec k_3)\cor{}{\vp_+(t,\vec k_j)\vp_-(t_*,-\vec k_j)}'\nonumber\\
&\phantom{=}+\Xi^{-;0}_{2,1}(\vec k_1,\vec k_2,\vec k_3)\cor{}{\vp_-(t_*,\vec k_j)\vp_+(t,-\vec k_j)}'\Big]\nonumber\\
&=\frac{a(t)^{-1}a^{\ve}_*}{8k_1k_2k_3}\sum_{j=1}^3k_j\Big[\Xi^{+;0}_{2,1}(\vec k_1,\vec k_2,\vec k_3)-\Xi^{-;0}_{2,1}(\vec k_1,\vec k_2,\vec k_3)\Big]\,.
\label{eq:F21explicit}
\end{flalign}

We now move on to the correlation functions in the second line of \eqref{eq:bimatch}. The insertion of the leading-power IC term \eqref{eq:F30} into the three-point functions containing one $\vp_-$-field gives
\begin{flalign}
&\Big[\cor{}{\vp_-(t,\vec k_1)\vp_+(t,\vec k_2)\vp_+(t,\vec k_3)}'+\textrm{permutations}\Big]\Big|_{\,\Xi_{3,0}}\nonumber\\
&=\frac{ia^{\ve}_*}{4k_1k_2k_3}\frac{i\pi\kappa_3}{2H}\sum_{j=1}^3k_j\Big[\cor{}{\vp_-(t,\vec k_j)\vp_+(t_*,-\vec k_j)}'+\cor{}{\vp_+(t_*,\vec k_j)\vp_-(t,-\vec k_j)}'\Big]=0\,,
\end{flalign}
where we used \eqref{eq:Xi30matching} to get the first equality, and \eqref{eq:vppm2pt} to get the second. Finally, we also have the contribution generated by the field redefinition \eqref{eq:cubicvpredef}. Since they are multiplied by the coupling $c_{3,0}$ we evaluate them in the Gaussian approximation, and find
\begin{flalign}
&(a(t)H)^{-1}\frac{c_{3,0}a(t)^{\ve}}{2\ve}\Big[\cor{}{\vp^2_+(t,\vec k_1)\vp_+(t,\vec k_2)\vp_+(t,\vec k_3)}'+\textrm{permutations}\Big]\Big|_{\textrm{free}}\nonumber\\
&=\frac{c_{3,0}}{4k_1k_2k_3}\frac{k_t}{a(t)H}\frac{a(t)^{\ve}}{\ve}\,.
\label{eq:biredef}
\end{flalign}

We now have all the pieces to match \eqref{eq:latebispectrum} at next-to-leading power. However, before doing so we need to renormalize the UV divergence we found in \eqref{eq:biredef}. Using \eqref{eq:bimatch} we find 
\begin{flalign}
&a(t)^{-3+3\ve}\hspace{-0.06cm}\bigg\{\hspace{-0.1cm}\cor{}{\vp_+(t,\vec k_1)\vp_+(t,\vec k_2)\vp_+(t,\vec k_3)}'_{|\,c_{2,1}+\Xi_{2,1}}\hspace{-0.2cm}+\hspace{-0.06cm}(a(t)H)^{-1}\hspace{-0.06cm}\bigg[\hspace{-0.04cm}\cor{}{\vp_-(t,\vec k_1)\vp_+(t,\vec k_2)\vp_+(t,\vec k_3)}'_{|\,\Xi_{3,0}}\nonumber\\
&+\frac{c_{3,0}a(t)^{\ve}}{2\ve}\cor{}{\vp^2_+(t,\vec k_1)\vp_+(t,\vec k_2)\vp_+(t,\vec k_3)}'_{|\,\textrm{free}}+\textrm{permutations}\bigg]\bigg\}\nonumber\\
&=\frac{a(t)^{-3+3\ve}}{4k_1k_2k_3}\frac{k_t}{a(t)H}\bigg\{c_{2,1}H+a(t)^{\ve}\bigg[\bigg(\frac{a_*}{a(t)}\bigg)^{\ve}\sum_{j=1}^3\frac{k_j}{2k_t}\Big[\Xi^{+;0}_{2,1}(\vec k_1,\vec k_2,\vec k_3)-\Xi^{-;0}_{2,1}(\vec k_1,\vec k_2,\vec k_3)\Big]\nonumber\\
&\phantom{=}+\frac{c_{3,0}}{H\ve}\bigg]\bigg\}\,.
\label{eq:bimatchcombination}
\end{flalign}
The divergence in the last line is renormalized by splitting the bare IC function $\Xi^{\sigma;0}_{2,1}$ into counterterm and finite part as
\begin{equation}
\Xi^{\sigma;0}_{2,1}(\vec k_1,\vec k_2,\vec k_3)=\xi^{\sigma}_{2,1}(\vec k_1,\vec k_2,\vec k_3)+\Xi^{\sigma}_{2,1}(\vec k_1,\vec k_2,\vec k_3)\,,
\end{equation}
and using the same CTP-index structure as in \cite{Beneke:2026rtf}:
\begin{equation}
\xi^+_{2,1}(\vec k_1,\vec k_2,\vec k_3)=-\xi^-_{2,1}(\vec k_1,\vec k_2,\vec k_3)=-\frac{c_{3,0}}{H\ve}\,.
\end{equation}
We can now take the limit $\ve\to0$ in \eqref{eq:bimatchcombination}, and match it to the next-to-leading power terms in \eqref{eq:latebispectrum}:
\begin{flalign}
&\frac{a(t)^{-3}}{4k_1k_2k_3}\frac{\kappa_3}{H}\frac{k_t}{a(t)H}\bigg[1-\log\bigg(\frac{e^{\gamma_E}k_t}{a(t)H}\bigg)\bigg]\nonumber\\
&=\frac{a(t)^{-3}}{4k_1k_2k_3}\frac{k_t}{a(t)H}\bigg[c_{2,1}H+\frac{H}{2}\sum_{j=1}^3\frac{k_j}{k_t}\Big[\Xi^{+}_{2,1}(\vec k_1,\vec k_2,\vec k_3)-\Xi^{-}_{2,1}(\vec k_1,\vec k_2,\vec k_3)\Big]\nonumber\\
&-c_{3,0}\log\bigg(\frac{a_*}{a(t)}\bigg)\bigg]\,.
\end{flalign}
Comparing the coefficients of the logarithm of the scale factor we match
\begin{equation}
c_{3,0}=\frac{\kappa_3}{H}\,,
\end{equation}
and the RPI relation \eqref{eq:coupRPI} then fixes
\begin{equation}
c_{2,1}=\frac{\kappa_3}{H^2}\,.
\end{equation}
Finally, the remainder of the logarithm can be reproduced by picking
\begin{equation}
\Xi^{+}_{2,1}(\vec k_1,\vec k_2,\vec k_3)=-\Xi^{-}_{2,1}(\vec k_1,\vec k_2,\vec k_3)=-\frac{\kappa_3}{H^2}\log\bigg(\frac{e^{\gamma_E}k_t}{a_*H}\bigg)\,,
\end{equation}
completing the matching at next-to-leading power. 

As anticipated, we find that in this case the tree-level matching computation of the bispectrum works according to the power-counting rules obtained from the free theory in \secref{sec:CCSdSET}. 

\section{Tree-level trispectrum in \texorpdfstring{$\phi^4$}{phi4}-theory and matching attempt}
\label{sec:trimatch}

We now move on to the CC $\phi^4$-theory in $d=4$ dimensions, and illustrate the consequence of its classical conformal invariance by computing the tree-level trispectrum
\begin{equation}
\cor{}{\phi(\eta,\vec k_1)\phi(\eta,\vec k_2)\phi(\eta,\vec k_3)\phi(\eta,\vec k_4)}\,,
\end{equation}
and attempting to match it onto the SdSET. We argue that in this case the necessary terms do not fit into the expected power-counting scheme, and trace the problem back to the (mis-)identification of the superhorizon degrees of freedom. 

\subsection{Full-theory computation}

The computation of interest here was already considered in \cite{Cohen:2020php}, and we repeat it using our notation and conventions. We need to evaluate
\begin{flalign}
&\cor{}{\phi(\eta,\vec k_1)\phi(\eta,\vec k_2)\phi(\eta,\vec k_3)\phi(\eta,\vec k_4)}'_{|\,\Lo(\kappa_4)}\nonumber\\
&=-\frac{\kappa_4(-H\eta)^{4}}{8k_1k_2k_3k_4}\Im\bigg[\int_{-\infty}^{\eta}\der\eta'\;e^{ik_t(\eta-\eta')}\bigg]\,.
\end{flalign}
Evaluating the above integral we find
\begin{equation}
\cor{}{\phi(\eta,\vec k_1)\phi(\eta,\vec k_2)\phi(\eta,\vec k_3)\phi(\eta,\vec k_4)}'_{|\,\Lo(\kappa_4)}=-\frac{\kappa_4(-H\eta)^4}{8k_1k_2k_3k_4k_t}\,,
\label{eq:treetri}
\end{equation}
where $k_i\equiv|\vec k_i|$ and $k_t=\sum_{i=1}^4k_i$. Up to the overall factor $(-H\eta)^4$ the result is fully time-independent. In the late-time limit $-k_i\eta\to0$ no simplification occurs, since the combination $-k_i\eta$ does not appear at all in \eqref{eq:treetri}. This is a direct consequence of the fact that, at tree level, we are effectively computing flat-space correlation functions of the massless field $\chi$ \eqref{eq:phichi}, which are insensitive to the expansion of the spacetime.

This may be contrasted with the example of the bispectrum \eqref{eq:latebispectrum}. In that case the cubic interaction breaks the conformal invariance already at tree level, and the resulting sensitivity of the theory to the dynamical spacetime manifests itself through the explicit dependence of the bispectrum on the combination $-k_t\eta$. As such, in that case the late-time limit $-k_t\eta\to0$ is non-trivial, while here it is. 

\subsection{Matching attempt}

As above, the starting point for the matching procedure is the matching equation between full-theory and SdSET four-point functions. To arrive at an effective-field basis with manifest power-counting we need to eliminate the interaction term $\sim c_{4,0}\vp^4_+$ from the SdSET action via the field redefinition
\begin{equation}
\vp_-(t,\vec x)\rightarrow\vp_-(t,\vec x)-\frac{a(t)^{-1+2\ve}}{1-2\ve}\frac{c_{4,0}}{3!}\vp^3_+(t,\vec x)\,,
\end{equation}
which follows from \eqref{eq:CCredef} with $n=4$. This introduces non-linear terms in $\vp_+$ in \eqref{eq:freephisplit} which, as in \secref{sec:bimatch}, are power-suppressed. For the present discussion it is sufficient to work at leading power in $\lambda$, so we may ignore them. Using \eqref{eq:freephisplit} the leading-power matching equation reads
\begin{flalign}
&\lim\limits_{k_i/(a(t)H)\rightarrow0}\cor{}{\phi(t,\vec k_1)\phi(t,\vec k_2)\phi(t,\vec k_3)\phi(t,\vec k_4)}\nonumber\\
&=\lim\limits_{\ve\to0}a(t)^{-4+4\ve}\cor{}{\vp_+(t,\vec k_1)\vp_+(t,\vec k_2)\vp_+(t,\vec k_3)\vp_+(t,\vec k_4)}\,.
\end{flalign}
By comparing powers of $a(t)$ multiplying the $\vp_+$ trispectrum with those appearing in \eqref{eq:treetri}, we see that they match. The result for the SdSET trispectrum therefore must be time-independent. The natural object to insert into the trispectrum that satisfies these properties is an IC term, which is also the conclusion that was drawn in \cite{Cohen:2020php}. The quartic IC functions enter via the terms
\begin{flalign}
&\sum_{\sigma=\pm}\int\bigg[\prod_{j=1}^4\frac{\der^{d-1}k_j}{(2\pi)^{d-1}}\bigg](2\pi)^{d-1}\delta^{(d-1)}\bigg(\sum_{l=1}^4\vec k_l\bigg)\sum_{m=0}^4\frac{a(t)^{-1-m}a^{2\ve}_*}{(4-m)!m!}\Xi^{\sigma;0}_{4-m,m}(\vec k_1,...,\vec k_4)\nonumber\\
&\times\prod_{l=m+1}^{4}\vp^{\sigma}_+(t_*,\vec k_l)\prod_{q=1}^m\vp^{\sigma}_-(t_*,\vec k_q)
\end{flalign}
in the IC functional \eqref{eq:FIC}. The insertion of these terms into the $\vp_+$ trispectrum leads to the following schematic result:
\begin{equation}
\cor{}{\vp_+(t,\vec k_1)\vp_+(t,\vec k_2)\vp_+(t,\vec k_3)\vp_+(t,\vec k_4)}'_{|\Xi_{4-m,m}}\sim \frac{a(t)^{-1-m}a^{2\ve}_*k^m_i}{k_1k_2k_3k_4}\Xi^{\sigma;0}_{4-m,m}(\vec k_1,...,\vec k_4)\,.
\end{equation}
Comparing the momentum- and $a(t)$-dependence of this result with \eqref{eq:treetri} fixes the following scaling for the IC function in question:
\begin{equation}
\Xi^{\sigma;0}_{4-m,m}(\vec k_1,...,\vec k_4)\sim\bigg(\frac{k_i}{a(t)}\bigg)^{-m-1}\,,\quad 0\leq m\leq 4\,.
\label{eq:ICproblem}
\end{equation}
This result is problematic, because, as remarked above, all terms in the SdSET action and IC functional should have manifest power-counting. In particular, this would imply that the quartic IC functions should scale like
\begin{equation}
\Xi^{\sigma;0}_{n-m,m}(\vec k_1,...,\vec k_n)\stackrel{!}{\sim}\bigg(\frac{k_i}{a(t)H}\bigg)^0\sim\lambda^0\,.
\end{equation}
However, we see from \eqref{eq:ICproblem} that $\Xi^{\sigma;0}_{4-m,m}\sim(H\lambda)^{-m-1}$. The coefficients of the operators in the IC functional are thus power-enhanced, contradicting the above statements. This signals that the late-time limit of the full-theory correlation function in question is not described correctly by the power-counting scheme of the EFT established above.

The problem in this case can be traced back to the identification of the interacting superhorizon degrees of freedom. The prescription leading to the power-counting \eqref{eq:CCpowercount} takes the free-theory mode functions as a starting point, assuming that they can be related to the interacting ones. When the interaction already breaks the conformal invariance of the free theory at tree level, such as in the $\phi^3$-case, this expectation holds, since the classical theory is already sensitive to the presence of the horizon of dS. The free-theory two-point function is the basic building block for the perturbative computation of correlation functions, so a perturbative relation between free and interacting effective degrees of freedom in this case may be constructed starting from tree-level matching. By contrast, in the CC $\phi^4$-theory the conformal invariance is broken only at loop level, and the superhorizon dynamics is entirely generated by quantum effects. 
This property obstructs the identification of the interacting late-time degrees of freedom taking the free-theory ones as a starting point. 

\section{Identification of leading superhorizon degrees of \texorpdfstring{\\}{} freedom in the \texorpdfstring{$\phi^4$}{phi4}-theory}
\label{sec:superhdofs}

We now address the question of how to correctly identify the leading superhorizon degrees of freedom of the CC $\phi^4$-theory. From the previous considerations it is clear that quantum corrections are the source of sensitivity of this theory to the presence of the horizon. Therefore, it follows that the effective superhorizon degrees of freedom are also generated by them. To identify them, we therefore wish to isolate their leading contribution. 

This can be done using the fact that any field in dS has the following leading behaviour in the late-time limit \cite{Arkani-Hamed:2015bza}:
\begin{equation}
\lim\limits_{\eta\rightarrow0}\phi(\eta,\vec x)=\sum_{\Delta}(-\eta)^{\Delta}O_{\Delta}(\vec x)\,,
\label{eq:latetsum}
\end{equation}
where $O_{\Delta}(\vec x)$ are primary operators of the Euclidean conformal field theory living on the future boundary of dS. Since the coupling in the theory under consideration is perturbative, in the limit $\kappa_4\rightarrow0$ the above sum continuously reduces to
\begin{equation}
\lim\limits_{\eta\rightarrow0}\phi_{\textrm{free}}(\eta,\vec x)=(-\eta)^{\Delta_+}O_{\Delta_+}(\vec x)+(-\eta)^{\Delta_-}O_{\Delta_-}(\vec x)\,,
\label{eq:philate}
\end{equation} 
with
\begin{equation}
\Delta_{\pm}=\frac{d-1}{2}\pm\frac{1}{2}
\end{equation}
the scaling dimensions of the free field, which coincide with $\alpha$ and $\beta$, see \eqref{eq:CCpowercount}. Therefore, the least-suppressed term in \eqref{eq:latetsum} should continuously go into $(-\eta)^{\Delta_-}O_-$ as $\kappa_4\rightarrow0$. In the interacting theory, we may assume that the $\Delta_{\pm}$ get corrected by a renormalization of the mass parameter $m^2$, and by the generation of a non-vanishing field anomalous dimension $\gamma_{\phi}$. Therefore, the scaling dimensions determining the late-time limit \eqref{eq:philate} of the interacting field read \cite{Bros:2010rku,Marolf:2010zp,Cohen:2024anu}
\begin{equation}
\Delta_{\pm}=\frac{d-1}{2}\pm\nu+\gamma_{\phi}\,,
\label{eq:scalingdimcor}
\end{equation}
where $\nu$ now contains the renormalized mass parameter. The two-point function of $\phi$ and its Fourier transform are related by
\begin{equation}
\cor{}{\phi(\eta,\vec x)\phi(\eta',\vec y)}=\int\frac{\der^{d-1}k}{(2\pi)^{d-1}}e^{i\vec k\cdot(\vec x-\vec y)}\cor{}{\phi(\eta,\vec k)\phi(\eta',-\vec k)}'
\end{equation}
exactly. Since the left-hand side is invariant under the dS isometries, and in particular it is invariant under dilatation transformations, the Fourier-transformed two-point function must transform under them as 
\begin{equation}
\cor{}{\phi(\eta,\vec k)\phi(\eta',-\vec k)}'\rightarrow\delta^{1-d}\cor{}{\phi(\eta,\vec k)\phi(\eta',-\vec k)}'\,,
\label{eq:latechi2pt}
\end{equation}
with $\delta$ the dilatation factor, to all orders in perturbation theory. Inspecting the free-theory result \eqref{eq:freeCC2pt}, we see that it already fulfills this property. Combining its functional form with \eqref{eq:scalingdimcor} we may conclude that the shift in the exponents of $\eta$ and $\eta'$ must be compensated by a corresponding shift in the momentum dependence of the two-point function, which then combines to the scale-invariant quantity $-k\eta^{(\prime)}$. This is then the leading $k$- and $\eta^{(\prime)}$-dependence of the $\phi$ two-point function which is sensitive to the presence of the horizon of dS. It can be most directly isolated by considering the rescaled field $\chi$. If the leading quantum correction to $\Delta_-$ only comes from $\gamma_{\phi}$ and no mass renormalization occurs, the perturbative shift of the scaling dimension of the fields due to the anomalous dimension modifies the late-time limit of the $\chi$ two-point function to 
\begin{equation}
\lim\limits_{\eta^{(\prime)}\rightarrow0}\cor{}{\chi(\eta,\vec k)\chi(\eta',-\vec k)}'\sim(-k\eta)^{\gamma_{\phi}}(-k\eta')^{\gamma_{\phi}}\frac{1}{k}\,,
\label{eq:chi2ptresummed}
\end{equation}
where we omitted a $k$- and $\eta^{(\prime)}$-independent proportionality constant.

\begin{figure}[t]
\centering
\begin{subfigure}{0.49\textwidth}
\raisebox{0.9cm}{\includegraphics[width=\textwidth]{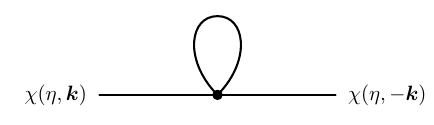}}
\caption{}
\end{subfigure}%
\begin{subfigure}{0.49\textwidth}
\includegraphics[width=\textwidth]{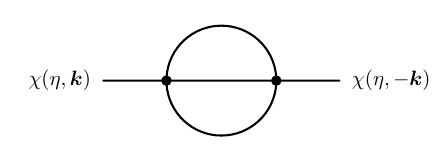}
\caption{}
\end{subfigure}
\caption{The one-loop tadpole (a) and two-loop sunset (b) diagrams.}
\label{fig:propcors}
\end{figure}

We now show that this is what occurs in the case under consideration. To do so, consider the action for the field $\chi$ \eqref{eq:chiS}, emphasizing that both the field and coupling are bare quantities by means of the subscript ``0", and allowing for a deviation from the strict conformal coupling to the metric due to the potential need for mass renormalization. The starting point is thus the bare $d$-dimensional action 
\begin{equation}
S=\int\der^dx\;\bigg[\frac{1}{2}\p_{\mu}\chi_0\p^{\mu}\chi_0-\frac{m^2_0}{2}a(\eta)^2\chi^2_0-\frac{\kappa_{4,0}}{4!}a(\eta)^{2\ve}\chi_0^4\bigg]\,.
\label{eq:ddimchiS}
\end{equation}
As before, in the above equation the indices of the derivatives are raised and lowered with the Minkowski metric $\eta_{\mu\nu}$. The bare objects need to be renormalized, as is standard, as
\begin{equation}
\chi_0=\sqrt{Z_{\phi}}\chi\,,\quad m^2_0=\delta m^2\,,\quad \kappa_{4,0}=\tmu^{2\ve}Z_{\kappa}\kappa_4\,,
\label{eq:rens}
\end{equation}
where we used that the field-strength-renormalization factors of $\phi$ and $\chi$ coincide. The counterterms $\delta_i$ are then defined as usual via $Z_i=1+\delta_i$. We wish to use dimreg with $d=4-2\ve$ to regularize the UV divergences, and we introduced the $\overline{\textrm{MS}}$ scale $\tmu=\sqrt{e^{\gamma_E}/(4\pi)}\mu$. We see from \eqref{eq:ddimchiS} and \eqref{eq:rens} that $a(\eta)$ and $\tmu$ appear in the action expressed in terms of renormalized quantities in the combination
\begin{equation}
(a(\eta)\tmu)^{2\ve}\,.
\end{equation}
This further suggests that the $\mu$-dependence and the ``anomalous" $a(\eta)$-dependence of the renormalized correlators are related to each other. 

The anomalous dimension $\gamma_{\phi}$ is computed from the field-strength renormalization constant $Z_{\phi}$ using the formula
\begin{equation}
\gamma_{\phi}\equiv\frac{1}{2}\frac{1}{Z_{\phi}}\frac{\der Z_{\phi}}{\der\log(\mu)}\,.
\label{eq:AD}
\end{equation}
Its leading expression in this case arises at the two-loop order, and it can be obtained by means of the following argument. The first correction to the two-point function of $\chi$ is due to the one-loop tadpole diagram shown on the left panel of \figref{fig:propcors}. However, this leads to an expression proportional to a scaleless, quadratically divergent momentum integral, which vanishes in dimreg:
\begin{flalign}
&\cor{}{\chi(\eta,\vec k)\chi(\eta,-\vec k)}'_{|\,\Lo(\kappa_4)}\nonumber\\
&=\frac{\tmu^{2\ve}\kappa_4}{8k^2}\Im\bigg[\int_{-\infty}^{\eta}\frac{\der\eta'}{(-H\eta')^{2\ve}}\;e^{2ik(\eta'-\eta)}\bigg]\int\frac{\der^{d-1}l}{(2\pi)^{d-1}}\frac{1}{l}=0\,.
\end{flalign}
The first non-vanishing correction then arises from the two-loop sunset diagram shown on the right panel of \figref{fig:propcors}. Up to $\ve$-dependent powers of $a(\eta)=(-H\eta)^{-1}$, the integrand is fully equivalent to its flat-space counterpart:
\begin{flalign}
&\cor{}{\chi(\eta,\vec k)\chi(\eta,-\vec k)}'_{|\,\Lo(\kappa^2_4)}\nonumber\\
&=-\frac{(\tmu^{2\ve}\kappa_4)^2}{192k^2}\int_{-\infty}^{\eta}\frac{\der\eta_1\der\eta_2}{(-H\eta_1)^{2\ve}(-H\eta_2)^{2\ve}}\int\frac{\der^{d-1}l_1\der^{d-1}l_2}{(2\pi)^{2d-2}}\frac{1}{l_1l_2|\vec l_1+\vec l_2-\vec k|}\nonumber\\
&\phantom{=}\times\bigg[\Big[e^{i[(2\eta-\eta_1-\eta_2)k+(l_1+l_2+|\vec l_1+\vec l_2-\vec k|)(\eta_1-\eta_2)]}+e^{-i[(2\eta-\eta_1-\eta_2)k+(l_1+l_2+|\vec l_1+\vec l_2-\vec k|)(\eta_1-\eta_2)]}\Big]\theta(\eta_1-\eta_2)\nonumber\\
&\phantom{=\times}\Big[e^{i[(\eta_1+\eta_2-2\eta)k+(l_1+l_2+|\vec l_1+\vec l_2-\vec k|)(\eta_1-\eta_2)]}+e^{-i[(\eta_1+\eta_2-2\eta)k+(l_1+l_2+|\vec l_1+\vec l_2-\vec k|)(\eta_1-\eta_2)]}\Big]\theta(\eta_2-\eta_1)\nonumber\\
&\phantom{=\times}-\Big[e^{i(\eta_1-\eta_2)(l_1+l_2+|\vec l_1+\vec l_2-\vec k|-k)}+e^{-i(\eta_1-\eta_2)(l_1+l_2+|\vec l_1+\vec l_2-\vec k|-k)}\Big]\bigg]\,.
\label{eq:twoloopchi}
\end{flalign}
The UV divergences arising from this diagram can be removed by means of the mass- and field-strength-renormalization counterterms $\delta m^2$ and $\delta_{\phi}$ introduced above, as is standard. Furthermore, since the first non-vanishing terms of these counterterms come from this diagram, they must be single poles in $\ve$. The flat-space version of the theory under consideration is properly regularized by dimreg, which implies that the time integrals in \eqref{eq:twoloopchi} require no separate regularization. Therefore, the presence of the factors $(-H\eta_{1,2})^{-2\ve}$ in \eqref{eq:twoloopchi} can only modify the finite parts of the result, but not the divergent part. Since all we are interested in is extracting the UV poles of this diagram, we may restrict ourselves to carrying out the flat-space computation instead, which amounts to deleting the factors $(-H\eta_{1,2})^{-2\ve}$ in \eqref{eq:twoloopchi}. Finally, the renormalization constants $Z_i$ of the field and coupling, as well as the mass counterterm $\delta m^2$, are universal, and do not depend on the type of observable being computed using the action in question. Therefore, instead of extracting the pole term from the resulting flat-space in-in correlation function, we may instead do so by computing its in-out counterpart, reducing the computation to a standard exercise. We thus compute the following amputated flat-space in-out correlation function:
\begin{equation}
-\frac{\kappa_4^2}{6}\int\frac{\der^dl_1}{(2\pi)^d}\int\frac{\der^dl_2}{(2\pi)^d}\frac{i^3}{[l^2_1+i\epsilon][l^2_2+i\epsilon][(l_1+l_2-p)^2+i\epsilon]}=\frac{i\kappa_4^2}{6144\pi^4\ve}p^2+\textrm{finite}\,.
\label{eq:flatsunset}
\end{equation}
All momenta appearing in the above equation are $d$-dimensional, and $p^2\equiv\eta_{\mu\nu}p^{\mu}p^{\nu}$. The dependence of the UV pole in \eqref{eq:flatsunset} on $p^2$ tells us that the appropriate counterterm to subtract it is
\begin{equation}
\delta_{\phi}=\kappa_4^2\bigg[-\frac{1}{6144\pi^4\ve}+\textrm{finite}\bigg]+\Lo(\kappa_4^3)\,,
\label{eq:phict}
\end{equation}
whereas no mass renormalization is needed at this order, so $\delta m^2$ is finite. For simplicity, we imagine that the mass parameter is determined at this perturbative order by some measurement, to still find the conformally-coupled value, such that $\delta m^2=\Lo(\kappa_4^3)$. The finite part of $\delta_{\phi}$ is renormalization-scheme dependent, but the pole part is not. While the momentum structure multiplying the UV pole resulting from the computation of \eqref{eq:twoloopchi} is certainly different, the above argument guarantees that the same pole part of $\delta_{\phi}$ \eqref{eq:phict} serves to subtract the UV divergence of that in-in correlator. Using \eqref{eq:AD} and the tree-level running of the renormalized $d$-dimensional coupling
\begin{equation}
\frac{\der\kappa_4}{\der\log(\mu)}=-2\ve\kappa_4+\Lo(\kappa^2_4)
\end{equation}
we then find the two-loop anomalous dimension
\begin{equation}
\gamma_{\phi}=\frac{\kappa^2_4}{3072\pi^4}+\Lo(\kappa^3_4)\,,
\label{eq:2loopAD}
\end{equation}
confirming the statement in \cite{Green:2020txs}. Combined with \eqref{eq:chi2ptresummed}, it yields the leading contribution to the late-time limit of the two-point function of $\chi$, and thus also of $\phi$, which is sensitive to the presence of the horizon of dS. Therefore, the form of the two-point function \eqref{eq:chi2ptresummed} should be used as a starting point to define the proper effective description for the superhorizon modes in this theory. Doing so is beyond the scope of this paper, and it is left for future work.

One particularly noteworthy aspect of the equal-time version of the result \eqref{eq:chi2ptresummed} is that its time- and renormalization-scale dependences coincide at the two-loop order:
\begin{equation}
\frac{\der}{\der\log(a(\eta))}\cor{}{\chi(\eta,\vec x)\chi(\eta,\vec y)}=\frac{\der}{\der\log(\mu)}\cor{}{\chi(\eta,\vec x)\chi(\eta,\vec y)}=-2\gamma_{\phi}\cor{}{\chi(\eta,\vec x)\chi(\eta,\vec y)}\,.
\end{equation}
This is a particular realization of the idea of the DRG \cite{Tanaka:1975ti,Boyanovsky:1998aa,Boyanovsky:2003ui,Burgess:2009bs} applied to correlation functions in dS. It was exploited in \cite{Green:2020txs} in a similar way as what we did here to achieve the resummation of secular logarithms of the form $\log(-k\eta)$ in the CC $\phi^4$-theory. Indeed, the expansion of the two-point function \eqref{eq:chi2ptresummed} with the two-loop anomalous dimension \eqref{eq:2loopAD} plugged in must generate all leading-logarithmic terms found in its explicit perturbative computation, to all orders in perturbation theory. Thus, in this specific example the standard RG resummation of logarithms of the form $\log(\mu/H)$ coincides with the resummation of the logarithms $\log(-k\eta^{(\prime)})$. However, this is due to the classical conformal invariance of the theory considered here, and it is not true in general. One of the motivations for the proposal of the SdSET in \cite{Cohen:2020php} was to systematize the idea of the DRG, thus achieving the resummation of secular logarithms, and it was applied in \cite{Cohen:2020php} to the CC $\phi^4$-theory as well, to recover the DRG resummation carried out in \cite{Green:2020txs}. We wish to stress here that the secular logarithms we are presently discussing are qualitatively different than in theories in dS displaying infrared divergences and secularly growing terms, such as the massless, minimally coupled $\phi^4$-theory considered in \cite{Cohen:2020php,Cohen:2021fzf,Beneke:2023wmt,Beneke:2026ksj,Beneke:2026rtf}, as well as the much less severe case of the CC $\phi^3$-theory considered in \secref{sec:bimatch}. Here, as well as in the examples considered in \cite{Green:2020txs}, the secular logarithms are generated by the UV divergences of the full theory. The resummation of these logarithms amounts to a perturbative shift of the leading time dependence of correlation functions at late times, through \eqref{eq:latetsum}, which is in turn related to the renormalization properties of the fields and parameters of the full theory. Therefore, in these theories such logarithms should be properly interpreted as \textit{ultraviolet} in nature, and not infrared. Consequently, their resummation should be carried out directly in the full theory, without the need for the SdSET. Above we argued that this resummation step in the full theory is necessary in order to properly identify the correct late-time degrees of freedom in this case. Thus, rather than being a result of the SdSET, in this case the resummation of these leading logarithms defines the starting point for the construction of the appropriate late-time EFT. By contrast, in theories which suffer from IR divergences and secular growth, these logarithms appear in the late-time limit already at tree level, see again \eqref{eq:latebispectrum} for a very mild example of this. These are genuine IR logarithms, which are not related to any UV poles of the full theory. They necessitate the use of the SdSET to control and eventually resum them, in order to obtain well-defined correlation functions of the full theory. In this case, these logarithms can be related to UV divergences of the \textit{effective} theory, which are not present in the full one.

\section{Conclusion}
\label{sec:conclusion}

In this work we examined the application of the SdSET to theories which enjoy a classical conformal invariance, using the conformally-coupled $\phi^4$-theory as an example. We pointed out the fact that the identification of the proper effective degrees of freedom in this class of theories is subtle, since the superhorizon dynamics is entirely generated by the quantum effects which break the conformal invariance of the classical theory.   
We showed explicitly that the standard SdSET construction and power-counting, based on the free conformally-coupled modes, succeeds in the case of the matching of the tree-level bispectrum of the conformally-coupled $\phi^3$-theory, where conformal invariance is already broken at tree level. By contrast, it fails to reproduce the tree-level trispectrum of the interacting $\phi^4$-theory within the expected power-counting scheme, which is signaled by the need for initial-condition functions with power-enhanced momentum dependence. This counterpoint illustrates that the obstruction encountered in this case is tied to the preservation of the classical conformal invariance by the quartic interaction term. To resolve this issue, we proposed that the proper effective degrees of freedom in such theories should be identified using the version of the two-point function of the field which includes its leading quantum correction as a starting point, since this is the first source of its sensitivity to the horizon of de Sitter. In this context, we further commented on the interpretation of secular logarithms of the form $\log(-k\eta)$ appearing in this theory, arguing that they should be properly interpreted as ultraviolet in nature, as they originate from the breaking of the conformal invariance due to renormalization. Thus, the resummation of the leading-logarithmic corrections to the two-point function constitutes the starting point for identifying the correct late-time degrees of freedom, and it is not the output of the SdSET. The aim of these results is to clarify an important feature of this EFT framework, which was not pointed out in previous works. They suggest that the effective late-time description of classically conformally-invariant theories in de Sitter space requires a different organization, which is intrinsically tied to the structure of quantum corrections to the theory. 

The theory discussed in this work shares these features with more realistic theories, such as four-dimensional gauge theories with massless fermions. It would thus be interesting to develop the structure of the SdSET for such theories, taking the observations in this work as a starting point, paving the way to its application to theories such as massless quantum electrodynamics and chromodynamics. As stressed above, the effective degrees of freedom of the EFT should be read off from the resummed two-point function of $\phi$, which is related to \eqref{eq:chi2ptresummed}. It is enough to include the two-loop result for $\gamma_{\phi}$ \eqref{eq:2loopAD} in the exponent of \eqref{eq:chi2ptresummed}, since higher-order corrections to it can be treated as small perturbations. Similarly, a non-vanishing perturbative correction to the mass parameter can be accommodated in \eqref{eq:chi2ptresummed}, which we did not consider in this paper for simplicity. The effective description of these loop-induced superhorizon degrees of freedom should then be matched onto resummed $n$-point functions. The resummation again only needs to take into account the leading quantum corrections to these objects, and the resulting matching procedure plays the role of the usual tree-level matching. Higher-order corrections to these leading expressions can then be treated perturbatively. These hypotheses should be checked using a theory in which the proposed resummation procedure is attainable, such as the $O(N)$ vector model in de Sitter space in the large-$N$ limit studied in \cite{DiPietro:2023inn}. Such an explicit construction would also shed light on the question of whether the resulting effective description is local. These interesting and non-trivial points are left for future work.

\subsubsection*{Acknowledgements}

We thank Jos\'e Santiago for discussions and feedback on all stages of the project, Martin Beneke, Mikael Chala, Tim Cohen, Patrick Hager, and Jos\'e Santiago for careful reading and comments on the manuscript, and Carlos Duaso Pueyo, Javier Fuentes, Harry Goodhew, Guilherme Guedes, and Kamran Salehi Vaziri for useful discussions. This work has been partially funded by the grants
EUR2024.153549, CNS2024-154834 funded by MCIN/AEI/10.13039/ 501100011033, PID2022-139466NB-C21 (FEDER/UE) funded by
the Spanish Research Agency (MICIU/AEI/10.13039/501100011033) and the European Union NextGeneration\\EU/PRTR.

\bibliography{Bibliography}{}
\end{document}